\documentclass[%
 reprint,
superscriptaddress,
 amsmath,amssymb,
aps,
]{revtex4-2}

\usepackage{graphicx}
\usepackage{dcolumn}
\usepackage{bm}
\usepackage{hyperref}
\usepackage{csquotes}

\usepackage[%showframe,%Uncomment any one of the following lines to test 
margin=0.5in,
]{geometry}

\begin{document}

\preprint{APS/123-QED}

\title{Super-resolution imaging of azimuthal features with illumination carrying OAM}

\author{Nilakshi Senapati}
\email{nilakshisenapati0408@gmail.com}
\author{Abhinandan Bhattacharjee}%
\altaffiliation[Currently at ]{Dept. of Physics, Paderborn University, Germany.}
\affiliation{%
	Department of Physics, Indian Institute of Technology Kanpur, Kanpur 208016, India
}
\author{Kedar Khare}
\affiliation{ Optics and Photonics Centre, Indian Institute of Technology Delhi, New Delhi 110016, India}
\author{Anand K Jha}
\email{akjha@iitk.ac.in}
\affiliation{%
 Department of Physics, Indian Institute of Technology Kanpur, Kanpur 208016, India
}%

\date{\today}

\begin{abstract}

Super-resolution imaging refers to imaging techniques that surpass the Rayleigh resolution limit. One standard way to achieve super-resolution is by structuring the phase of the field illuminating the object. Although super-resolution techniques are already employed in commercial imaging devices, intense research efforts continue to enhance the resolution even further. In this work, we show that if the field illuminating the object is structured in the azimuthal coordinate--such as a field carrying orbital angular momentum (OAM)--the azimuthal features of the object can be imaged with enhanced imaging resolution. We experimentally demonstrate it with two objects, namely, an azimuthal double-slit and a Siemens star. We find that for a given azimuthal feature, there is an optimum OAM mode index of the illumination that gives the best imaging resolution. Super-resolution imaging of azimuthal feature can have important implications, especially for some biological objects that are known to have predominantly azimuthal features.

\end{abstract}

\maketitle

The maximum imaging resolution that an imaging system can achieve is known as the Rayleigh resolution criterion. Enhancing  
resolution beyond the Rayleigh resolution limit is referred to as super-resolution, and it has been a very active area of research in the last few decades with invention of techniques such as stimulated emission depletion microscopy (STED) \cite{hell1994optlett, klar1999optlett}, structured illumination microscopy (SIM) \cite{heintzmann2002josaa, gustafsson2005pnas}, spatial-mode demultiplexing (SPADE) \cite{rouviere2024optica, tsang2016prx}, and Fourier Ptychyography (FT) \cite{zheng2013natphot}. Among these techniques, STED  \cite{hell1994optlett, klar1999optlett} and SIM \cite{heintzmann2002josaa, gustafsson2005pnas} use spatially coherent illumination but collect incoherent fluorescence. SPADE\cite{rouviere2024optica, tsang2016prx} uses incoherent illumination but collects spatially-coherent light modes, while in FT \cite{zheng2013natphot} the illumination and collection involve coherent light. Another very active area of research in recent decades has been light fields carrying orbital angular momentum (OAM) \cite{allen1992pra, padgett2017optexp}. Such fields have emerged as powerful tools in areas such as long-distance communication \cite{wang2012natphot, suciu2023pra, yan2014natcomm, bozinovic2013science}, increased error tolerance in quantum communication \cite{cerf2002prl, nikolopoulos2006pra}, gate implementation \cite{ralph2007pra, lanyon2009natphy}, propagation through turbulence \cite{bhattacharjee2022sciadv}, quantum metrology \cite{jha2011pra} and fundamental tests of quantum mechanics \cite{kaszlikowski2000prl, collins2002prl, vertesi2010prl, dada2011natphy}.

Light fields carrying OAM have also been used for edge imaging \cite{furhapter2005optexp, zeng2022jlt} and for fluorescence imaging such as STED microscopy \cite{hell1994optlett, klar1999optlett}. Furthermore, it has been shown that when light coming from two spatially separated point sources is made to go through an OAM imparting forked hologram,  one can resolve the two sources at angular separation one order of magnitude below the Rayleigh criterion \cite{tamburini2006prl}.  However, the use of OAM modes for direct super-resolution imaging has been very limited and most of the works have been related to theoretical modelling and numerical simulations \cite{li2013pre, yang2021ijrs, ashrafian2025scirep, liang2022pre}. These works hint at super-resolution-imaging potential of fields carrying OAM modes but do not point out the unique and the optimum use of such fields for enhancing the imaging resolution. Furthermore, to the best of our knowledge, there has been no experimental demonstration in direct imaging of super-resolution with illumination carrying OAM.

In contrast, in this work, we report experimental demonstrations of super-resolution imaging with illumination carrying OAM. We find that the unique benefit of illumination carrying OAM is in super-resolving the azimuthal features of an object and that if the object has predominantly non-azimuthal features, the OAM modes  are not the optimum illumination for achieving super-resolution. We also derive the mathematical relation for the optimal OAM mode for a given azimuthal feature.

\begin{figure*}%[t!]
	\centering
	\includegraphics[width=0.95\textwidth]{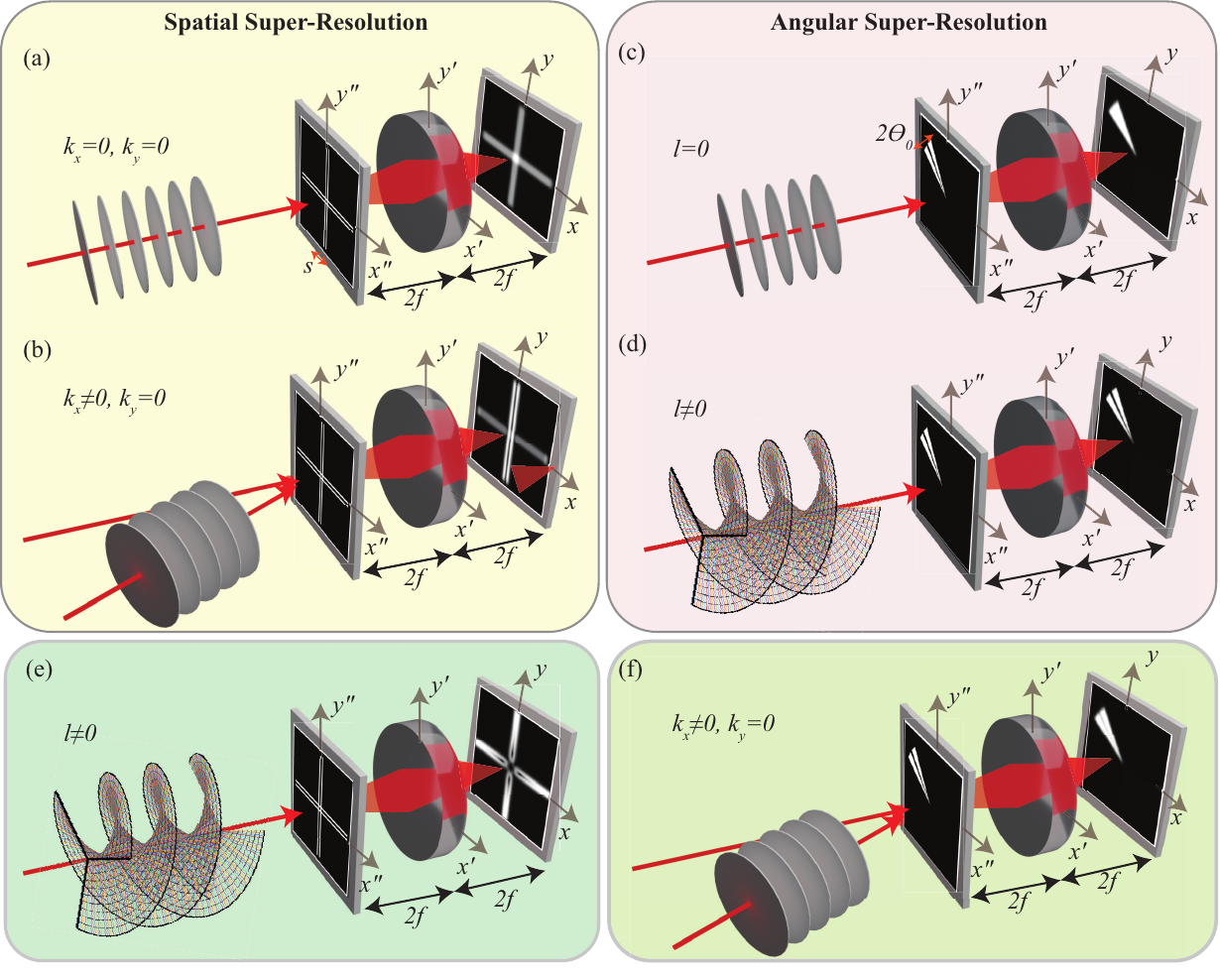}
	\caption{Illustrating how structuring the wavefront of the illumination in a particular coordinate helps super-resolve the object features in that coordinate. The object present in (a), (b), and (e) is in the form of a transverse double-slit, with two slits in both $x$- and $y$-directions. The slit-width and slit-separation of the object is $0.0625$ mm and $0.125$ mm, respectively. The object present in (c), (d), and (f) is an azimuthal double-slit, with the angular slit width being $0.02\pi$ and the angular slit separation being $0.1\pi$. The focal length of the lens is $400$ mm. For each sub-figure, we have explicitly shown the transmission function of the object at $z=0$ and the corresponding image-plane intensity at $z=4f$ calculated numerically using Eq.~(\ref{I4f}).}
	\label{fig1}
\end{figure*}

Figure \ref{fig1} shows imaging configurations with various objects and illuminations. We use these to illustrate how structuring the wavefront of the illumination affect imaging resolution. In every configuration of Fig.~\ref{fig1}, the object is placed at $z=0$, the imaging lens at $z=2f$, and the image plane is located at $z=4f$, where $f$ is the focal length of the lens. The electric field at $z=0$ plane is given by $E(x'',y'';z=0)=E_{\rm in}(x'',y'';z=0)T(x'',y'')$, where $E_{\rm in}(x'',y'';z=0)$ is the illumination field and $T(x'',y'')$ is the object transmission function. This field undergoes Fresnel propagation from $z=0$ plane to the lens plane ($z=2f$), where it acquires the quadratic phase due to the lens and then further propagates to the image plane. Therefore, the electric field at $z=4f$ becomes (see Supplementary Information Sec.~I.A and Sec.~5.3 of \cite{goodman2005})
\begin{align*}
&E(x,y;z=4f) = e^{-\frac{ik}{4f}(x^{2}+y^{2})} \iint E_{\rm in}(x'',y'',z=0) \\ &\times T(x'',y'') e^{\frac{ik}{4f}(x''^{2}+y''^{2})} e^{-\frac{k^{2}d^{2}}{4f^{2}}(x+x'')^{2}}  e^{-\frac{k^{2}d^{2}}{4f^{2}}(y+y'')^{2}} dx'' dy'', 
\end{align*}
where $d$ is the radius of the lens-aperture. The image-plane intensity $I(x,y,z=4f) = |E(x,y,z=4f)|^{2}$  then becomes 
\begin{multline}
I(x,y,z=4f) = e^{-\frac{x^{2}+y^{2}}{\sigma_{p}^{2}}} \biggl|  \iint E_{\rm in}(x'',y'',z=0) T(x'',y'') \\ \exp\biggl[-\frac{(x''^{2}+y''^{2})}{2\sigma_{p}^{2}} \biggl( 1-i\frac{k\sigma_{p}^{2}}{2f} \biggr) \biggr] \exp\biggl[ -\frac{x x'' + y y''}{\sigma_{p}^{2}} \biggr] dx'' dy'' \biggr|^{2}, \label{I4f}
\end{multline}
where $\sigma_{p} = \frac{2f}{kd}$.  The objects present in the left column of sub-figures in Fig.~\ref{fig1} is in the form of a transverse double-slit, with a double-slit in both $x$- and $y$-directions. The object present in the right column of sub-figures is an azimuthal double-slit. For the transverse-double-slit object, we take the slit width and slit separation to be $0.0625$ mm and $0.125$ mm, respectively. For the azimuthal-double-slit object, we take the angular width to be $0.02\pi$ and the angular slit separation to be $0.1\pi$.  For each sub-figure in Fig.~\ref{fig1}, we have explicitly shown the transmission function of the object at the $z=0$ plane and the corresponding image-plane intensity calculated numerically using Eq.~(\ref{I4f}) at the $z=4f$ plane. For all the numerical calculations, we have taken $d=2.5$ mm, $f=400$ mm, and $\lambda=633$ nm. We note that the values of the various object and illumination parameters have been chosen such that the object features cannot be resolved under the plane-wave illumination---this is to highlight the effect of structured illumination on enhancing image resolution that we next demonstrate. We show that structuring the illumination in a particular coordinate leads to super-resolution of the object-features in that coordinate.

\begin{figure*}%[t!]
	\centering
	\includegraphics[width=0.95\textwidth]{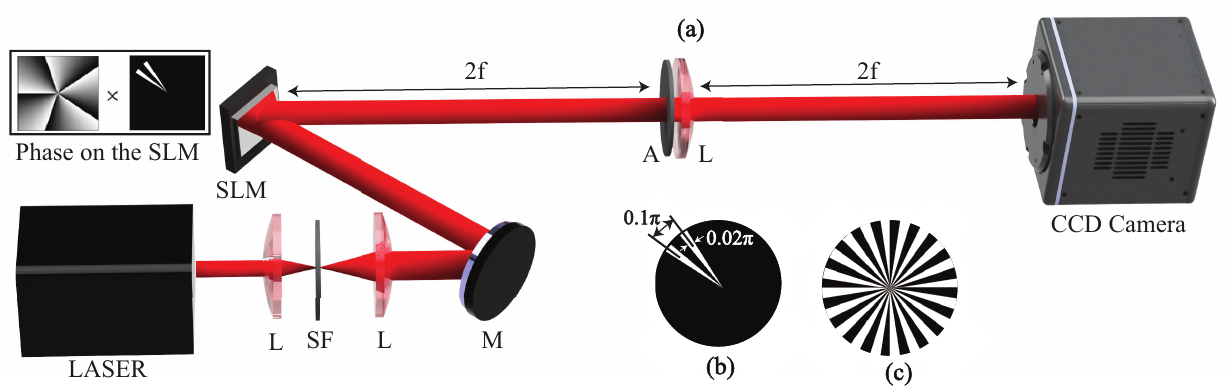}
	\caption{(a) Schematic of the experimental setup. (b) The azimuthal-double-slit object. (c) The Siemens star object. For the objects shown in (b) and (c), the angular slit width is $0.02\pi$ and the angular slit separation is $0.1\pi$. SF: spatial filter, SLM: spatial light modulator, A: aperture, L: lens, M: mirror.}
	\label{fig2}
\end{figure*}

Figures~\ref{fig1}(a) and \ref{fig1}(b) depict imaging of transverse-double-slit object when the illumination is by a plane wave $(k_x=k_y=0)$ and a plane wave tilted in $x$-direction $(k_x\neq0; k_y=0)$, respectively. We find that the plane-wave illumination is not able to resolve the object. However, the tilted plane wave is able to super-resolve the two slits in the $x$-direction but not in the $y$-direction. This is because the phase of titled plane wave changes as $k_x x$, which structures the phase in the $x$-direction but not in the $y$-direction, leading to super-resolution only in the $x$-direction. It can be shown that one obtains optimum super-resolution when the phase of the illuminating field at the two slit locations differs by $\pi$ \cite{marcet2008optlett, liang2010optlett, ferreira2015annphy, pietersoone2024spie} (see Supplementary Information Sec.~I.D for for an intuitive explanation). This result can be extended to derive a relation between the slit separation $s$ and the optimum $k_x$ for super-resolution. The relationship is given by $k_x s=(2n+1)\pi$ (see Supplementary Information Sec.~I.B for more details). Figure~\ref{fig1}(c) depicts imaging of the azimuthal-double-slit object illuminated by a field carrying no OAM $(l=0)$ while in Fig.~\ref{fig1}(d) the illuminating field carries finite OAM $(l\neq 0)$. We find that while the $l=0$ illumination is not able to resolve the object, the illumination carrying non-zero OAM leads to super-resolution. Again, this is because the phase of an OAM carrying  is structured as  $l\theta$, and it therefore leads to super-resolution of azimuthal features. 

Figure \ref{fig1}(e) depicts imaging of the transverse-double-slit object illuminated with a field carrying OAM. Although the resolution in the case is better than that with the plane wave, it is worse compared to resolution with the tilted plane wave depicted in Fig.~\ref{fig1}(c). Similarly, Fig.~\ref{fig1}(f) depicts imaging of the azimuthal-double-slit object illuminated with the tilted plane wave. Again, in this case, although the resolution is better than that with beam carrying no OAM, it is worse compared to the resolution with OAM carrying field, as depicted in Fig.~\ref{fig1}(d). Thus for direct imaging, structuring the illumination does not lead to super-resolution of every object feature. For the optimum super-resolution imaging of a given object feature, the illumination needs to be structured accordingly. This fact could be utilized in proposals \cite{li2013pre, yang2021ijrs, ashrafian2025scirep, liang2022pre} in which OAM carrying illumination is employed for super-resolution of every object features. If the light is structured in $x$-direction  it can lead to optimal resolution enhancement only in the $x$-direction. Similarly, for optimal super-resolution of the azimuthal features, the illumination needs to be structured in the azimuthal direction.

\begin{figure*}%[t!]
	\centering
	\includegraphics[width=0.95\textwidth]{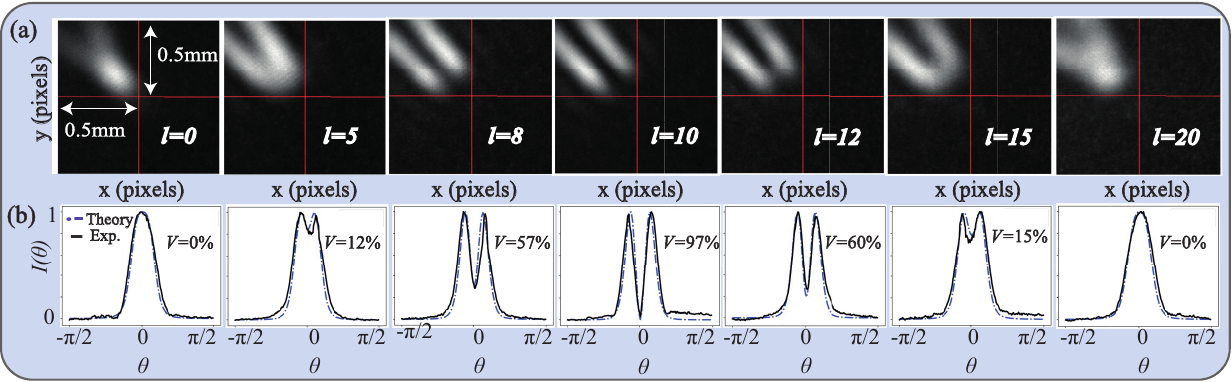}
	\caption{Experimental results. (a) Image intensity $I(r, \theta, z=4f)$ of the azimuthal-double-slit object shown in Fig.~\ref{fig2}(b) for various OAM mode index $l$ of the illumination. (b) The plot of the radially averaged intensity $I(\theta)$, defined in Eq.~(\ref{avg-Int}), as a function of $\theta$ for various plots shown in (a). The visibility $V$ of the intensity pattern $I(\theta)$ is indicated on each plot. }
	\label{fig3}
\end{figure*}
\begin{figure*}%[h!]
	\centering
	\includegraphics[width=0.95\textwidth]{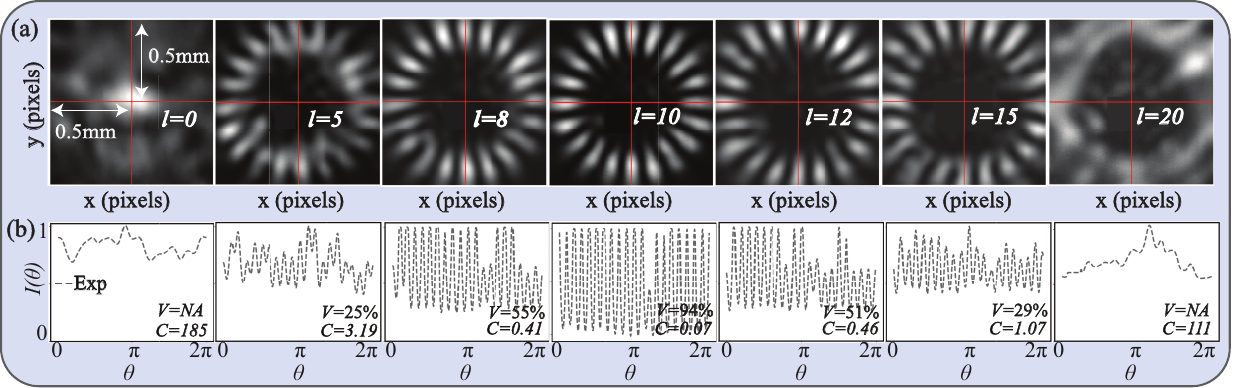}
	\caption{Experimental results. (a) Image intensity $I(r, \theta, z=4f)$ of the Siemens star object shown in Fig.~\ref{fig2}(c) for various OAM mode index $l$ of the illumination. (b) The plot of the radially averaged intensity $I(\theta)$, defined in Eq.~(\ref{avg-Int}), as a function of $\theta$ for various plots shown in (a). The visibility $V$ of the intensity pattern $I(\theta)$ and the Cramér–Rao Lower Bound $C$ are indicated on each plot.} \label{fig4}
\end{figure*}
So far, through the numerical simulations of Eq.~(\ref{I4f}), we have shown that the fields carrying OAM lead to super-resolution of azimuthal features. In what follows, we derive a relation for the optimum modes $l_{\rm opt}$ that best resolves an azimuthal feature, namely, the angular separation in an azimuthal-double-slit. To this end, we express the image-plane intensity of Eq.~(\ref{I4f}) in the polar coordinates by using the transformation: $(x,y)=(r\cos\theta,r\sin\theta)$ and $(x'',y'')=(r''\cos\theta'', r''\sin\theta'')$. We take $E_{\rm in}(x'', y'', z=0)\rightarrow E_{\rm in}(r'',\theta'', z=0)=\exp\bigl[-\frac{r''^{2}}{4w^{2}}\bigr]e^{il\theta''}$ where $w$ is the beam-waist of the field and $l$ is the OAM mode index. The object is taken to have only the azimuthal feature, that is, $T(r'',\theta'')=T(\theta'')$. Eq.~(\ref{I4f}) thus becomes: 
\begin{align}
I(r,\theta,z=4f) &= A e^{\bigl(-\frac{r^{2}}{\sigma^{2}_{p}}\bigr)} \biggl| \iint T(\theta'') e^{il\theta''} \exp{\biggl[-\frac{a r''^{2}}{2\sigma_{p}^{2}}\biggr]} \notag\\ &\times \exp{\biggl[-\frac{r'' r}{\sigma_{p}^{2}}cos(\theta-\theta'')\biggr]} r'' dr'' d\theta'' \biggr|^{2}, \label{IOAM}
\end{align}
where $a = 1 + \frac{\sigma_{p}^{2}}{2w^{2}} - i \frac{k\sigma_{p}^{2}}{2f}$.
Next, we consider an azimuthal-double-slit object of infinitesimally narrow slit width and angular separation of $2\theta_0$ such that the transmission function is given by $T(\theta'')=\delta(\theta''-\theta_{0})+\delta(\theta''+\theta_{0})$. For this object, the image-plane intensity $I(r,\theta,z=4f)$ can be expressed as
\begin{align}
&I(r, \theta,z=4f)= A e^{-\frac{r^{2}}{\sigma^{2}_{p}}} \biggl| \iint \left[\delta(\theta''-\theta_{0})+\delta(\theta''+\theta_{0})\right] \notag \\
&\times e^{il\theta''} \exp{\left[-\frac{a r''^{2}}{2\sigma_{p}^{2}}\right]}\exp{\left[-\frac{r'' r}{\sigma_{p}^{2}}\cos(\theta-\theta'')\right]} r'' dr'' d\theta'' \biggr|^{2}.\notag \\
&= \left|f(r, \theta-\theta_0)e^{il\theta_0}+f(r, \theta+\theta_0)e^{-il\theta_0}\right|^2\label{IOAM1}
\end{align}
where
\begin{align}
f(r, \theta)=A e^{-\frac{r^{2}}{\sigma^{2}_{p}}} \int \exp{\left[-\frac{a r''^{2}}{2\sigma_{p}^{2}}\right]}\exp{\left[-\frac{r'' r}{\sigma_{p}^{2}}\cos\theta\right]} r'' dr''.
\end{align}
Now, at $\theta=0$, that is, at the mid-point between the two slits, the intensity $I(r,\theta,z=4f)$ is given by $
I(r,\theta=0,z=4f) = |f(r, -\theta_0)|^2  + |f(r, \theta_0)|^2 + 2f(r, -\theta_0)f(r, \theta_0)\cos 2l\theta_0.$ Here, we have assumed the function $f(r, \theta)$ to be real. Now, taking $f(r, -\theta_0)=f(r, \theta_0)=f_0$, we obtain 
\begin{align}
I(r,\theta=0,z=4f) = 2|f_0|^2 (1+\cos 2l\theta_0).\label{resolution}
\end{align}
For the maximum image contrast, we have to have $I(r,\theta=0,z=4f)=0$. This condition is satisfied when $\cos 2l\theta_0=-1$, which gives the optimum OAM mode index $l_{\rm opt}$ at which the contrast is maximum:
\begin{align}
l_{\rm opt}=\frac{(2n+1)\pi}{2\theta_0}, \label{IOAM4}
\end{align}
where $n=0, \pm1, \pm2, \cdots$. (see Supplementary Information Sec.~I.C for more details of the above calculations.) The smallest OAM mode index at which the contrast becomes maximum is given by $\frac{\pi}{2\theta_0}$. We note that when the illuminating field is a plane-wave $(l=0)$, instead of an OAM carrying field, we obtain $I(r,\theta=0,z=4f)=4|f_0|^2$. The contrast and hence resolution in this case is much worse---this has been illustrated through Fig.~\ref{fig1}(c) and \ref{fig1}(d). We further note that if the illumination is azimuthally incoherent, instated of being coherent, the cosine term in Eq.~(\ref{resolution}) becomes zero giving $I(r,\theta=0,z=4f)=2|f_0|^2$.  The resolution with incoherent light defines the Rayleigh resolution limit. Thus we see that the imaging contrast with a plane wave is worse than the Rayleigh resolution limit while with an OAM carrying field it surpasses the Rayleigh limit.

Figure \ref{fig2}(a) shows the experimental setup. A Helium Neon (He-Ne) laser at $633$ nm is used as a light source. The laser is spatially filtered in order to generate a high-quality Gaussian beam which is then made to fall onto the spatial light modulator (SLM) to generate fields with different OAM mode index  by displaying the corresponding holograms on the SLM \cite{arrizon2007josaa}. The SLM is kept at $z=0$, the lens at $z=2f$ and the CCD camera at $z=4f$. In addition to generate the illumination, the SLM also holds the objects. The objects are  displayed on the SLM and then imaged onto the multi-pixel CCD camera using a lens of focal length $f=400$ mm in a $1:1$ conjugate. The aperture-size of the lens is kept at $d=2.5\mathrm{mm}$ to keep the numerical aperture of the imaging system such that the plane-wave illumination is not able to resolve the objects. We use two separate objects with only azimuthal features. The first one is an azimuthal double-slit as shown in Fig.~\ref{fig2}(b) and the other one is the Siemens star as shown in Fig.~\ref{fig2}(c). For both the objects, the angular slit-width is $0.02\pi$ and the angular slit separation is $0.1\pi$.  We note that in our experiments, we use Laguerre-Gaussian modes. Although these modes have $e^{il\theta}$ dependence, the intensity distribution is non-uniform in the radial direction with a null in the center of the beam. In order to produce fields with nearly-uniform intensity in the radial direction and with $e^{il\theta}$ dependence, we mix Laguerre-Gaussian modes with varying beam-waists with a fixed OAM mode index $l$.

For the object in Fig.~\ref{fig2}(b), the measured image intensities are presented in Fig.~\ref{fig3}(a) for various OAM modes index $l$ of the illumination. To quantify the resolution enhancement, we use visibility of images in the following manner. First, using the expression in Eq.~(\ref{I4f}), we calculate the radially-averaged intensity given by: 
\begin{align} 
I(\theta) = \int r I(r,\theta,z=4f) dr. \label{avg-Int}
\end{align} 
We then define the visibility as $V = \frac{I(\theta)_{\text{max}} - I(\theta)_{\text{min}}}{I(\theta)_{\text{max}} + I(\theta)_{\text{min}}}$. We measure $V$ as a function of $l$ and indicate that on each plot in Fig.~\ref{fig3}(b). We find that field with $l=0$ is not able to resolve the azimuthal-double-slit. However, the resolution improves with increasing $l$, reaching the maximum visibility at $l_{\rm opt} = 10$, verifying the relation given in Eq.~(\ref{IOAM4}) for $2\theta_0=0.1\pi$. When the OAM is increased beyond $l=10$, the visibility decreases reaching its minimum value at $l = 20$ and then returns to its maximum at $l=30$.

For the object in Fig.~\ref{fig2}(c), the measured image intensities are presented in Fig.~\ref{fig4}(a) for various OAM modes index $l$ of the illumination. We measure $V$ as a function of $l$ and present the results in Fig.~\ref{fig4}(b). We find that unlike the plots in Fig.~\ref{fig3}(b), these plots have more experimental noise. Therefore, for quantifying the azimuthal resolution in this case, in addition to measuring the visibility, we also evaluate the Fisher Information (FI) and take the corresponding Cramér–Rao Lower Bound (CRLB) $C$ \cite{barbieri2022prxquant, motka2016epj} as a suitable metric (see Supplementary Information Sec.~II for more details). We note that a smaller value for $C$ implies better imaging resolution. Both $V$ and $C$ are indicated on each plot in Fig.~\ref{fig4}(b). We clearly see that $C$ is lowest for $l=10$, indicating optimum super-resolution and verifying the relation given in Eq.~(\ref{IOAM4}) for $2\theta_0=0.1\pi$.

In summary, in this work, we show that if the field illuminating the object is structured in the azimuthal coordinate--such as a field carrying orbital angular momentum (OAM)--the azimuthal features of an object can be imaged with enhanced resolution. We experimentally demonstrate it with two objects, namely, an azimuthal double-slit and a Siemens star. We find that for a given azimuthal feature, there is an optimal OAM mode index of the illumination that gives the best resolution. Being able to image azimuthal features with enhanced resolution can have important implications for object with predominantly azimuthal features, such as the centrosome in animal cells \cite{lawo2012nature}.

\vspace{0.01mm}
\paragraph*{Data Availability:} The data are available from the authors upon reasonable request.

\paragraph*{Acknowledgements:} N.S. thanks the Prime Minister’s Research Fellowship (PMRF), Government of India, for financial support. We acknowledge financial support from the
Science and Engineering Research Board through grants STR/2021/000035 and CRG/2022/003070 and from the Department of Science and Technology, Government of India through grant DST/ICPS/QuST/Theme-­ 1/2019 and through the National Quantum Mission (NQM).

\nocite{*}

\bibliography{angular_superresolution_imaging}

\section*{Supplemental Material: Super-resolution imaging of azimuthal features with illumination carrying OAM}

This document contains supplementary information in support of the manuscript \enquote{Super-resolution imaging of azimuthal features with illumination carrying OAM}.

\section{\label{math}Mathematical Derivation} \label{basics}
The detail mathematical derivation in support of the main paper is presented here. This supplemental material is organised in two parts, the mathematical derivation in Sec.~\ref{basics} and quantification in Sec.~\ref{quantification}. Inside the Sec.~\ref{basics}, at first we discuss about the basics of imaging \ref{imaging}, following the mathematics behind the resolution enhancement in spatial \ref{spatial} and angle-OAM degree of freedom \ref{angular}, then connect it with a nice physical intuitive model in the discussion \ref{discussion}. 

\subsection{Basics of Imaging} \label{imaging}
\begin{figure*}[t!]
	\centering
	\includegraphics[width=0.88\textwidth]{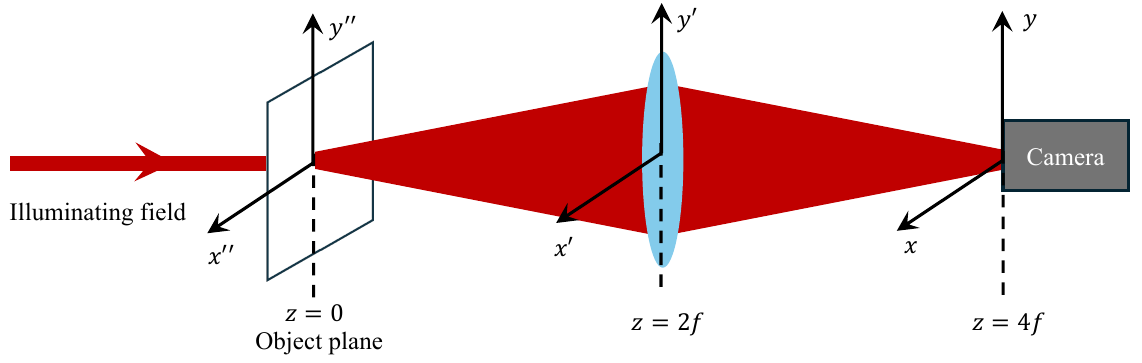}
	\caption{Schematic of a setup to image an object using a $2f-2f$ imaging system.}
	\label{fig1supp}
\end{figure*}

We start our discussion with the basic mathematical formulation behind a simple imaging system. We consider $(x'',y'')$ is the object plane, $(x',y')$ is the lens plane and $(x,y)$ is the image plane as shown in the Fig.~\ref{fig1supp}. We begin by defining the electric field at the object plane as $E(x'',y'';z=0) = E_{in}(x'',y'',z=0) T(x'',y'')$, where, $E_{in}(x'',y'',z=0)$  represents illuminating field and $T(x'',y'')$ is the object transmission function. Next, we employ Fresnel propagation to describe the evolution of the field from the object plane to a distance of $z=2f$ at $(x',y')$ plane. It is given by,
\begin{multline}
	E(x',y';z=2f) = e^{-\frac{ik}{4f}(x'^{2}+y'^{2})} \iint E(x'',y'';z=0) \\ \times e^{-\frac{ik}{4f}(x''^{2}+y''^{2})} e^{\frac{ik}{2f}(x' x'' + y' y'')} dx'' dy''.
\end{multline}
Consider the lens transmission function $T_{L} = e^{-\frac{(x'^{2}+y'^{2})}{2d^{2}}} e^{\frac{ik}{2f}(x'^{2}+y'^{2})}$ where $d$ is the effective aperture size, and $f$ is the focal length of the lens respectively. The electric field amplitude just after the lens is given by
\begin{equation}
	E_{lens}(x',y';z=2f) = e^{-\frac{(x'^{2}+y'^{2})}{2d^{2}}} e^{\frac{ik}{2f}(x'^{2}+y'^{2})} E(x',y';z=2f).
\end{equation}
We then apply Fresnel propagation from the lens plane to the image plane at $z=4f$ and we can write it as,
\begin{multline}
	E(x,y;z=4f) = e^{-\frac{ik}{4f}(x^{2}+y^{2})} \iint E_{lens}(x',y';z=2f) \\ \times e^{-\frac{ik}{4f}(x'^{2}+y'^{2})} e^{\frac{ik}{2f}(x x' + y y')} dx' dy' ,
\end{multline}
\begin{align}
	E(x,y;z=4f)= e^{-\frac{ik}{4f}(x^{2}+y^{2})} \iint E_{in}(x'',y'',z=0) T(x'',y'')  \notag \\ \times e^{-\frac{ik}{4f}(x''^{2}+y''^{2})}  \Biggl[ \int e^{-\frac{x'^{2}}{2d^{2}} + \frac{ik}{2f}(x+x'')x'} dx' \Biggr] \notag \\ \times \Biggl[ \int e^{-\frac{y'^{2}}{2d^{2}} + \frac{ik}{2f}(y+y'')y'} dy' \Biggr] dx'' dy''.
\end{align}
Now we use the standard gaussian integration formula $
\int e^{-\alpha x^{2} + \beta x} dx = \biggl( \frac{\pi}{\alpha} \biggr)^{1/2} e^{\bigl( \frac{\beta^{2}}{4\alpha} \bigr)}$ to solve the two integral inside, which becomes,
\begin{multline}
	\hspace{1cm}\int e^{-\frac{x'^{2}}{2d^{2}} + \frac{ik}{2f}(x+x'')x'} dx'  \approx e^{-\frac{d^{2}k^{2}}{4f^{2}} (x+x'')^{2}} \\ \text{and} \quad
	\hspace{2mm}\int e^{-\frac{y'^{2}}{2d^{2}} + \frac{ik}{2f}(y+y'')y'} dy'  \approx  e^{-\frac{d^{2}k^{2}}{4f^{2}} (y+y'')^{2}}.
\end{multline}
After substituting and performing the required standard integrations, we simplify the equation to obtain the electric field at the image plane as,
\begin{multline}
	E(x,y;z=4f) = e^{-\frac{ik}{4f}(x^{2}+y^{2})} \iint E_{in}(x'',y'',z=0) \\ \times T(x'',y'') e^{\frac{ik}{4f}(x''^{2}+y''^{2})} e^{-\frac{k^{2}d^{2}}{4f^{2}}(x+x'')^{2}}  e^{-\frac{k^{2}d^{2}}{4f^{2}}(y+y'')^{2}} dx'' dy''. \label{finalE}
\end{multline}
We write the intensity distribution as $	I_{l}(x,y,z=4f) = |E(x,y,z=4f)|^{2}$, and substituting $\frac{2f}{kd}$ as $\sigma_{p}$, which leads to
\begin{multline}
	I_{l}(x,y,z=4f) = \\ e^{-\frac{x^{2}+y^{2}}{\sigma_{p}^{2}}} \biggl|  \iint E_{in}(x'',y'',z=0) T(x'',y'') \times \\ \exp\biggl[-\frac{(x''^{2}+y''^{2})}{2\sigma_{p}^{2}} \biggl( 1-i\frac{k\sigma_{p}^{2}}{2f} \biggr) \biggr] exp\biggl[ -\frac{x x'' + y y''}{\sigma_{p}^{2}} \biggr] dx'' dy'' \biggr|^{2}. \label{finalintensity}
\end{multline}
\subsection{Resolution in Spatial degree of freedom} \label{spatial}
\begin{figure*}[t!]
	\centering
	\includegraphics[width=0.58\textwidth]{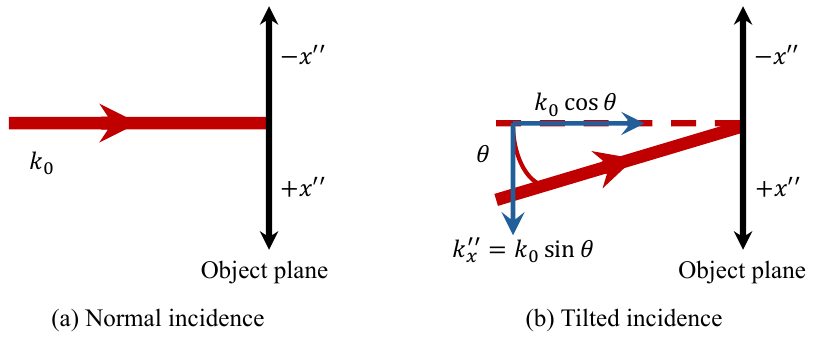}
	\caption{Schematic representation of the tilting geometry.}
	\label{fig2supp}
\end{figure*}
The input field is a gaussian plane wave, which can be written as 
$E_{in}(x'',y'')=e^{i(k_{x}''x''+k_{y}''y'')}e^{-\frac{(x''^{2}+y''^{2})}{2 w^{2}}}$. For the simplest case, considering the object as two dirac delta like slits separated by a distance $s$ in $x$-direction only i.e., $T(x'',y'')=T(x'')=\delta(x''-\frac{s}{2})+\delta(x''+\frac{s}{2})$. By substituting $E_{in}(x'',y'')$ and $T(x'',y'')$ in Eq.~\ref{finalintensity} and writing $\bigl( 1 + \frac{\sigma_{p}^{2}}{2w^{2}} - i \frac{k\sigma_{p}^{2}}{2f} \bigr)$ as $a$, then we get, 
\begin{equation}
	I(x,y)_{z=4f} =  \biggl| f\bigl(x-\frac{s}{2},y\bigr) e^{ik''_{x}s/2}+f(x+\frac{s}{2},y)e^{-ik''_{x}s/2}\biggr|^{2}
\end{equation}
\begin{subequations}
	\begin{equation}
		\text{where,}\quad f\bigl(x-\frac{s}{2},y\bigr) \\= e^{-\frac{s^{2}a}{8\sigma^{2}_{p}}} e^{-xs/2} \int e^{-\frac{y''^{2}a}{2\sigma^{2}_{p}}} e^{-\frac{yy''}{\sigma^{2}_{p}}} e^{ik_{y}y''} dy''
	\end{equation}
	\begin{equation}
		\text{and}\quad f\bigl(x+\frac{s}{2},y\bigr) \\= e^{-\frac{s^{2}a}{8\sigma^{2}_{p}}} e^{xs/2} \int e^{-\frac{y''^{2}a}{2\sigma^{2}_{p}}} e^{-\frac{yy''}{\sigma^{2}_{p}}} e^{-ik_{y}y''} dy''
	\end{equation}
\end{subequations}
Now at $x=0$, that is at the mid-point between the two slits, the intensity $I(x,y)_{z=4f}$ becomes
\begin{align}
	I(x=0,y)_{z=4f} = |f(-\frac{s}{2},y)|^{2} + |f(\frac{s}{2},y)|^{2} \notag \\ + 2 f(-\frac{s}{2},y) f(\frac{s}{2},y) \cos(k''_{x}s).
\end{align}
Here we have assumed the function to be real, thus we can write $f(-\frac{s}{2},y)=f(\frac{s}{2},y)=f_{0}$. Hence we obtain
\begin{equation}
	I(x=0,y)_{z=4f} = 2 |f_{0}|^{2} (1 + \cos(k''_{x}s)).
\end{equation}
When $cos(k''_{x}s)=-1$, then the intensity exactly vanishes at the midpoint between the slits. This situtation corresponds to the point of maximum resolution. Extending this argument further we can express it as follows:
\begin{equation}
	 k_{x}''s=(2n+1)\pi \implies s k_{0}\sin\theta = (2n+1)\pi, \label{spatialfinal}
\end{equation}
where, $\theta$ is the angle of the incident wave w.r.t. the normal (see Fig.~\ref{fig2supp}). This formula has a practical limitation. As the angle $\theta$ increases, the tilt of the incident wavefront becomes more pronounced. However, beyond a certain threshold of $\theta$, the corresponding shift in Fourier space grows excessively, leading to image distortion. 
If we consider an object with both $x-y$ features and calculate the above formula, then there will be another term corresponding to the $k_{y}''$. And if tilting in one direction enhances the resolution in particular direction (say $x$) then it will be compensated with the resolution degradation in another direction. So, tilting does not magically increase resolution. If we collect multiple tilted illuminations and combine them, then overall resolution enhancement emerges. This fundamental concept is actually being used for structured illumination microscopy(SIM) \cite{heintzmann2002josaa, gustafsson2005pnas}, Fourier ptychography (FT) \cite{zheng2013natphot}, synthetic aperture/off-axis illumination \cite{streibl1985josaa}. With Eq.~\ref{spatialfinal} we represent a simple way of understanding how tilting a plane wave can help in resolution enhancement in a particular direction. For a practical three-dimensional case it is very intuitive in the $k-$space. If we consider a sphere with three axes $k_{x}, k_{y}, k_{z}$ then $|k|$ is the radius and for a particular wavelegth beam this $|k|$ is constant. Then basically tilting corresponds to moving along the surface of this sphere.
\subsection{Resolution in angle-OAM degree of freedom}\label{angular}

Here we detail the mathematical derivation to describe the reasons of getting better visibility using OAM and cause of $100\%$ visibility using the similar toy model as it is used for the spatial case above (this part is already discussed in the main paper). We consider that the object only has the azimuthal features, that is $T(r'',\theta'')=T(\theta'')$. For this, we consider an azimuthal-double-slit object of infinitesimally narrow slit width and angular separation of $2\theta_{0}$ such that the transmission function is given by $T(\theta'')= [\delta(\theta''-\theta_{0})+\delta(\theta''+\theta_{0})].$
Now if we write \ref{finalintensity} in polar coordinate, then it becomes
\begin{align}
	I(r,\theta,z=4f) = A e^{\bigl(-\frac{r^{2}}{\sigma^{2}_{p}}\bigr)} \biggl| \iint T(\theta'') e^{il\theta''} \exp{\biggl[-\frac{a r''^{2}}{2\sigma_{p}^{2}}\biggr]} \notag \\ \times \exp{\biggl[-\frac{r'' r}{\sigma_{p}^{2}}\cos(\theta-\theta'')\biggr]} r'' dr'' d\theta'' \biggr|^{2}, \label{IOAM}
\end{align}
where again $\bigl( 1 + \frac{\sigma_{p}^{2}}{2w^{2}} - i \frac{k\sigma_{p}^{2}}{2f} \bigr)$ is denoted as $a$. The image-plane intensity $I(r,\theta,z=4f)$ can be expressed as
\begin{align}
	I(r, \theta,z=4f)= A e^{-\frac{r^{2}}{\sigma^{2}_{p}}} \biggl| \iint \left[\delta(\theta''-\theta_{0})+\delta(\theta''+\theta_{0})\right] \notag \\
	\times e^{il\theta''} \exp{\left[-\frac{a r''^{2}}{2\sigma_{p}^{2}}\right]}\exp{\left[-\frac{r'' r}{\sigma_{p}^{2}}\cos(\theta-\theta'')\right]} r'' dr'' d\theta'' \biggr|^{2}.
\end{align}
After applying the dirac delta function operation we obtain
\begin{equation}
	I(r, \theta,z=4f) = \left|f(r, \theta-\theta_0)e^{il\theta_0}+f(r, \theta+\theta_0)e^{-il\theta_0}\right|^2; \label{If}
\end{equation}
\begin{align}
	\text{where} \quad f(r, \theta)=A e^{-\frac{r^{2}}{\sigma^{2}_{p}}} \int \exp{\left[-\frac{a r''^{2}}{2\sigma_{p}^{2}}\right]} \notag \\ \times \exp{\left[-\frac{r'' r}{\sigma_{p}^{2}}\cos\theta\right]} r'' dr''.
\end{align}
Now, at $\theta=0$, that is, at the mid-point between the two slits, \ref{If} can be written as:
\begin{multline*}
	I(r,\theta=0,z=4f)\\ = |f(r, -\theta_0)|^2  + |f(r, \theta_0)|^2 + f(r, -\theta_0)f(r, \theta_0)(e^{i2l\theta_{0}}+e^{-i2l\theta_{0}}), \\
	= |f(r, -\theta_0)|^2  + |f(r, \theta_0)|^2 + 2f(r, -\theta_0)f(r, \theta_0)\cos 2l\theta_0.
\end{multline*}
Here, we have assumed the function $f(r, \theta)$ to be real, thus we write $f(r, -\theta_0)=f(r, \theta_0)=f_0$. Thus the above equation is written as 
\begin{align}
	I(r,\theta=0,z=4f) = 2|f_0|^2 (1+\cos 2l\theta_0). \label{resolution}
\end{align}
For maximum image contrast, the intensity at the midpoint must vanish, i.e.,$I(r,\theta = 0, z = 4f) = 0.$ This condition is satisfied when $\cos(2l\theta_0) = -1,$ which leads to the condition deduced below:
\begin{equation*}
	cos(2 l \theta_{0})=-1 \implies 2 l \theta_{0}=(2n+1)\pi \implies  2 l \theta_{0}= (2n+1)\pi.
\end{equation*}
Finally it determines the optimum OAM mode index \(l_{\rm opt}\) that yields the highest contrast:
\begin{align}
	l_{\rm opt}=\frac{(2n+1)\pi}{2\theta_0}. \label{IOAM4}
\end{align}
Here, $2\theta_{0}$ is the angular separation between the two angular slits. If we consider $2\theta_{0}=\alpha$, then it can be rewritten as $l\alpha = (2n+1)\pi$.
\subsection{Discussion}\label{discussion}

\begin{figure*}[t!]
	\centering
	\includegraphics[width=0.98\textwidth]{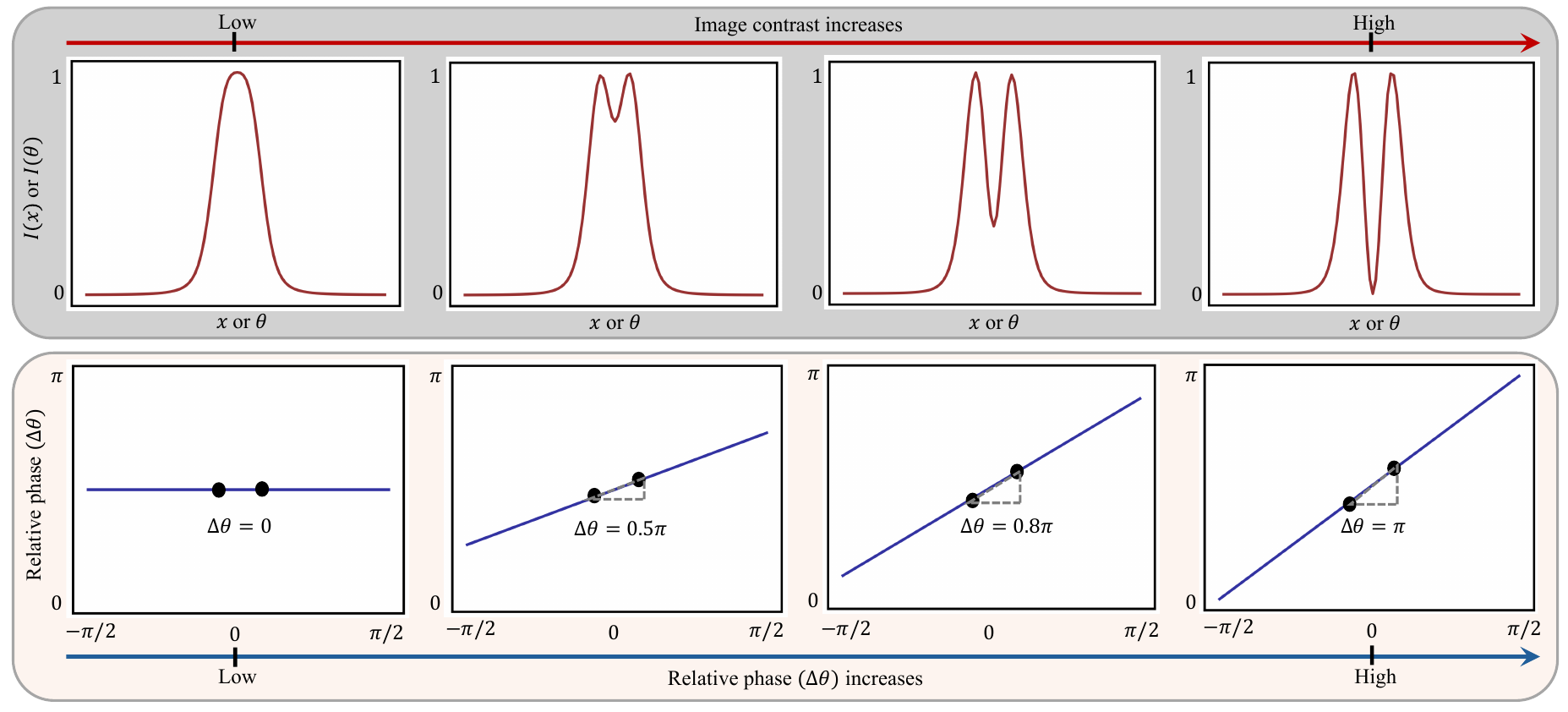}
	\caption{Visualization of physical understanding which shows that with the relative phase image contrast increases whereas it reaches to the highest resolution when relative phase is $\pi$. }
	\label{fig3supp}
\end{figure*}

By comparing Eq.~\ref{spatialfinal} from spatial imaging with Eq.~\ref{IOAM4} angular imaging, an direct analogy can be drawn: higher tilting provide better resolution in spatial degree of freedom for a spatially structured oject, similarly high OAM ($l$) mode provides better resolution while imaging an azimuthally structured object. This introduce the concept of naturally optimal beam choices tailored to specific object structures. For a transverse double slit, increasing the relative phase difference between the slits from 0 to $\pi$ enhances image contrast, peaking at $\pi$ \cite{goodman1985} as the midpoint falls to zero. Similarly, for two azimuthal slits, an OAM beam with an azimuthally varying phase can be tuned to create a $\pi$ phase difference, giving the highest contrast. This physical interpretetion is presented in a pictorial form in Fig.~\ref{fig3supp}. 
However, in the spatial case, controlling the tilt of the coherent illumination to reveal such periodicity is not practical, since beyond a certain angle the Fourier frequency shift becomes abrupt, ultimately degrading resolution and distorting the image. In contrast, because $l$ directly governs phase structuring, OAM beams naturally exhibit this periodic resolution enhancement in a well-controlled manner. Therefore, in case of spatial degree of freedom there is a practical limitation which we can refer as band-limited imaging. Also, here we consider the simplest case and for one dimension only, that's why Eq.~\ref{spatialfinal} looks as a clean formula but in realistic case we need to consider both dimensions as well as the practical limit of propagation wave. Thus, in practice, a direct or literal application of Eq.~\ref{spatialfinal} is not particularly suitable, rather SIM \cite{heintzmann2002josaa, gustafsson2005pnas}, FT \cite{zheng2013natphot}, these are much more applicable. Interestingly, this limitation can be overcome in the angular–OAM degree of freedom, owing to its inherent phase structuring.

\section{\label{quant}Qantification: Fisher Information and Cram{\'e}r--Rao Lower Bound}\label{quantification}

For objects exhibiting complex angular features, visibility by itself is insufficient as a measure of image quality. A more rigorous assessment can be obtained through the evaluation of Fisher Information (FI) and the associated Cramér–Rao Lower Bound (CRLB) \cite{barbieri2022prxquant, motka2016epj}. Since the intensity distribution at the image plane acts as a probability density function (PDF), i.e., $p(\theta \mid \theta_0)=I(\theta, z=4f)$, where $\theta_0$ represents the angular separation to be estimated, we define FI ($F(\theta_{0})$) and CRLB ($\Delta$) as follows:
\begin{align}
	F(\theta_0) &= \int \left[ \frac{\partial}{\partial \theta_0} \log p(\theta \mid \theta_0) \right]^2 p(\theta \mid \theta_0) d\theta, \\
	\Delta &\geq \frac{1}{F(\theta_0)}.
\end{align}
For objects like the Siemens star, where determining $I_{\text{max}}$ and $I_{\text{min}}$ is ambiguous, due to both experimental noise and complexity of the object. Thus visibility is not a suitable metric. Therefor, we also compute FI and CRLB, which provide a more meaningful quantification. We observe that the minimum CRLB corresponding the lowest variance in estimating $\theta_0$ occurs for $l = 10$, which aligns with our theoretical model. Here we roughly compute FI by doing single parameter estimation but for more precise quantification two-parameter or multi-parameter estimation would be more suitable. 

\nocite{*}

\end{document}